\PassOptionsToPackage{unicode}{hyperref}
\PassOptionsToPackage{hyphens}{url}
\PassOptionsToPackage{dvipsnames,svgnames,x11names}{xcolor}
\documentclass[a4paper,11pt]{article}
\usepackage[margin=2.5cm]{geometry}
\usepackage{float}
\usepackage{caption}
\usepackage{orcidlink}
\usepackage{xcolor}
\usepackage{amsmath,amssymb}
\usepackage{iftex}
\ifPDFTeX
  \usepackage[T1]{fontenc}
  \usepackage[utf8]{inputenc}
\else 
  \usepackage{unicode-math} 
  \defaultfontfeatures{Scale=MatchLowercase}
  \defaultfontfeatures[\rmfamily]{Ligatures=TeX,Scale=1}
\fi
\usepackage[]{stix}
\ifPDFTeX\else
\fi
\IfFileExists{upquote.sty}{\usepackage{upquote}}{}
\IfFileExists{microtype.sty}{
  \usepackage[]{microtype}
  \UseMicrotypeSet[protrusion]{basicmath} 
}{}
\makeatletter
\ifx\paragraph\undefined\else
  \let\oldparagraph\paragraph
  \renewcommand{\paragraph}{
    \@ifstar
      \xxxParagraphStar
      \xxxParagraphNoStar
  }
  \newcommand{\xxxParagraphStar}[1]{\oldparagraph*{#1}\mbox{}}
  \newcommand{\xxxParagraphNoStar}[1]{\oldparagraph{#1}\mbox{}}
\fi
\ifx\subparagraph\undefined\else
  \let\oldsubparagraph\subparagraph
  \renewcommand{\subparagraph}{
    \@ifstar
      \xxxSubParagraphStar
      \xxxSubParagraphNoStar
  }
  \newcommand{\xxxSubParagraphStar}[1]{\oldsubparagraph*{#1}\mbox{}}
  \newcommand{\xxxSubParagraphNoStar}[1]{\oldsubparagraph{#1}\mbox{}}
\fi
\makeatother

\usepackage{longtable,booktabs,array}
\usepackage{calc} 
\usepackage{etoolbox}
\makeatletter
\patchcmd\longtable{\par}{\if@noskipsec\mbox{}\fi\par}{}{}
\makeatother
\IfFileExists{footnotehyper.sty}{\usepackage{footnotehyper}}{\usepackage{footnote}}
\makesavenoteenv{longtable}
\usepackage{graphicx}
\makeatletter
\newsavebox\pandoc@box
\newcommand*\pandocbounded[1]{
  \sbox\pandoc@box{#1}%
  \Gscale@div\@tempa{\textheight}{\dimexpr\ht\pandoc@box+\dp\pandoc@box\relax}%
  \Gscale@div\@tempb{\linewidth}{\wd\pandoc@box}%
  \ifdim\@tempb\p@<\@tempa\p@\let\@tempa\@tempb\fi
  \ifdim\@tempa\p@<\p@\scalebox{\@tempa}{\usebox\pandoc@box}%
  \else\usebox{\pandoc@box}%
  \fi%
}
\def\fps@figure{htbp}
\makeatother

\NewDocumentCommand\citeproctext{}{}

\makeatletter
 \let\@cite@ofmt\@firstofone
 \def\@biblabel#1{}
 \def\@cite#1#2{{#1\if@tempswa , #2\fi}}
\makeatother
\newlength{\cslhangindent}
\newlength{\csllabelwidth}
\newenvironment{CSLReferences}[2] 
 {\begin{list}{}{%
  \setlength{\itemindent}{0pt}
  \setlength{\leftmargin}{0pt}
  \setlength{\parsep}{0pt}
  \ifodd #1
   \setlength{\leftmargin}{\cslhangindent}
   \setlength{\itemindent}{-1\cslhangindent}
  \fi
  \setlength{\itemsep}{#2\baselineskip}}}
 {\end{list}}
\usepackage{calc}

\newcommand{\CSLLeftMargin}[1]{\parbox[t]{\csllabelwidth}{\strut#1\strut}}
\newcommand{\CSLRightInline}[1]{\parbox[t]{\linewidth - \csllabelwidth}{\strut#1\strut}}

\providecommand{\tightlist}{%
  \setlength{\itemsep}{0pt}\setlength{\parskip}{0pt}}

\usepackage{bookmark}
\IfFileExists{xurl.sty}{\usepackage{xurl}}{} 
\makeatletter
\@ifundefined{xmpquote}{\newcommand{\xmpquote}[1]{#1}}{}
\makeatother
\hypersetup{
  pdftitle={A Physically Informed Changepoint Score for Fault Prioritisation in District Heating},
  pdfauthor={Tanja Korsten Bugajski; Nicolai Peder Bülow Pedersen; J. Eduardo Vera-Valdés; Michal Zbigniew Pomianowski; Anna Marszal-Pomianowska},
  pdfkeywords={\xmpquote{District heating}, \xmpquote{Smart heat
meters}, \xmpquote{Changepoint detection}, \xmpquote{Bayesian
inference}, \xmpquote{Fault prioritisation}, \xmpquote{Predictive
maintenance}},
  colorlinks=true,
  linkcolor={blue},
  filecolor={Maroon},
  citecolor={Blue},
  urlcolor={Blue},
  pdfcreator={LaTeX via pandoc}}

\begin{document}

\title{A Physically Informed Changepoint Score for Fault Prioritisation in District Heating}

\author{%
Tanja Korsten Bugajski\textsuperscript{1}\orcidlink{0009-0001-5546-8116},
Nicolai Peder Bülow Pedersen\textsuperscript{1}\orcidlink{0009-0006-9405-4828},\\
J. Eduardo Vera-Valdés\textsuperscript{1,3,}\thanks{Corresponding author: \texttt{eduardo@math.aau.dk}}\ \orcidlink{0000-0002-0337-8055},
Michal Zbigniew Pomianowski\textsuperscript{2}\orcidlink{0000-0002-9187-201X},\\
Anna Marszal-Pomianowska\textsuperscript{2}\orcidlink{0000-0002-3195-7388}\\[1ex]
\small \textsuperscript{1}Department of Mathematical Sciences, Aalborg University, Thomas Manns Vej 23, 9220 Aalborg Øst, Denmark\\
\small \textsuperscript{2}Department of the Built Environment, Aalborg University, Thomas Manns Vej 23, 9220 Aalborg Øst, Denmark\\
\small \textsuperscript{3}CoRE, Aarhus University, Aarhus, Denmark
}
\date{}
\maketitle

\begin{abstract}
Poorly cooling district-heating installations raise return temperatures
and pumped flow, increasing heat losses and pumping energy and hindering
low-temperature operation. Faults often go unreported for weeks, so
their onset is unknown and energy is wasted in the meantime. Utilities
flag poorly cooling installations by a low supply--return temperature
difference, but not when the problem began. Changepoint detectors such
as the Bayesian Estimator of Abrupt Change, Seasonality, and Trend
(BEAST) can date a change but report only its probability, so a large,
fault-like change and a marginal one can look alike. We therefore
develop a decision score that weighs each BEAST change in temperature
difference and flow by its probability, size, deviation from recent
operation and physical direction. Joint falls in temperature difference
and rises in flow, the signature of insufficient cooling, rank first.
Across 1,522 cases, 648 with a reported fault, 24 of the 25
highest-ranked cases were reported faults, and Average Precision was
0.772 against a prevalence of 0.426, likely conservative, as unreported
cases may contain faults. Without using the report date, the strongest
flow rise in reported-fault cases came a median of 11 days before the
report (128 days in comparison cases). Combined with
temperature-difference screening, the score raised Average Precision in
season-matched cases from 0.486 to 0.646. After the detected change,
flagged reported-fault installations returned water 16 K warmer and
about half their circulated water delivered no heat. The score gives
utilities a short list for inspection and a candidate onset date for
each change.
\end{abstract}

\medskip
\noindent\textbf{Keywords:} District heating; Smart heat meters; Changepoint detection; Bayesian inference; Fault prioritisation; Predictive maintenance

\newpage{}

\section{Introduction}\label{introduction}

Low return temperatures are central to the efficiency of district
heating. When a customer installation cools the supplied water poorly,
its return temperature rises and more water must be circulated to
deliver the same heat. The network then incurs higher heat losses and
pumping energy, and heat production by combined heat and power plants,
heat pumps and flue-gas condensation becomes less efficient (1,2). Low
return temperatures are also a precondition for the transition to
fourth-generation, low-temperature district heating (2). Detecting
poorly cooling installations early is therefore an energy-efficiency
measure as much as a maintenance task.

Smart heat meter data provide increasingly detailed information about
the operation of district-heating substations and create opportunities
for automated monitoring, fault detection, and predictive maintenance
(3). However, deviations in household-level time series are not
necessarily indicative of faults. Changes may arise from sensor or
communication problems, altered demand, for example due to changes in
occupancy (4), control behaviour, maintenance interventions, weather
conditions, or ordinary operating variability. Effective fault detection
therefore requires methods that are sufficiently sensitive to identify
abnormal behaviour while remaining interpretable in terms of the
physical operation of the heating system (1,5,6).

Previous work has approached this problem using statistical screening,
rule-based methods, unsupervised anomaly detection, machine learning,
and fault-diagnosis frameworks (7--10). More recent research has
increasingly emphasised the operational consequences of detected
abnormalities, including fault handling, maintenance prioritisation, and
predictive maintenance (11--14). A persistent challenge across these
approaches is the availability and quality of ground-truth fault
information. Operational service records may contain inaccurate
classifications, may omit faults that were never reported, and often
record the time of customer contact or maintenance rather than the
actual onset of abnormal behaviour (9,15). Consequently, a detected
anomaly cannot generally be matched to a reported fault at an exact
observation time, making event-level assessment more appropriate than
pointwise classification in many operational settings (9,14,16).

The timing problem is more fundamental than a matter of measurement
error. A fault report is, by necessity, created after the fault has
appeared, and the delay between onset and report may be days or weeks.
Even when a report exists, the date on which the fault actually began is
therefore not known. This matters for any attempt to learn the type of
fault from data. A classifier trained on data labelled by report date
alone mislabels observations in both directions: a window ending at the
report and assigned to a fault class will include measurements from
before the onset, when the installation was operating normally, while
data treated as normal because it precedes the report may already
include the early stages of the fault. Both errors blur the distinction
between fault types and between faulty and healthy operation. Training a
fault-type classifier thus requires that each fault be dated properly,
so that only the faulty part of the record is assigned to its class
(9,15).

Leiria et al.~examined this problem using fault reports and smart heat
meter data from Danish residential substations (15). Their detailed
analysis showed substantial variation in the manifestation of reported
faults and highlighted return temperature as a particularly consistent
indicator of abnormal operation, whereas energy-consumption patterns
were considerably less informative. They used self-organising maps
combined with k-means clustering to characterise fault symptoms and
operating patterns. To move beyond manual inspection and enable analysis
at larger scale, they further proposed automated time-series
decomposition and segmentation, specifically identifying the Bayesian
Estimator of Abrupt Change, Seasonality, and Trend (BEAST) as a
promising method for detecting abrupt changes, level shifts,
degradation, and other anomalous temporal behaviour. BEAST was proposed
as a direction for automated fault-detection workflows rather than
implemented and evaluated as an operational fault-detection method in
that study.

This motivates the present work. We investigate BEAST as a probabilistic
changepoint detector for household heat meter data (17). BEAST
represents uncertainty in the number and locations of structural changes
and provides posterior evidence for individual changepoints. In the
present analysis, the seasonal component is disabled and BEAST is used
to identify candidate changes in the underlying trend structure.
Detected changepoints also localise abrupt changes in time, so they
provide candidate onset dates that do not depend on the report date.
Importantly, these changepoints are not interpreted as faults directly.
A statistically well-supported structural change may represent either
deterioration or improvement in operation, or may arise from changes
unrelated to a fault.

BEAST's output by itself tells only how probable it is that a change
occurred. A large jump and a much smaller one can receive the same
posterior probability, and a list of such probabilities does not help a
utility decide which installation to inspect first, since the
installation with the severe, fault-like change cannot be told apart
from the one with a marginal change. Prioritisation requires a score
that also reflects how large the change is, how far it departs from the
installation's own recent operation, and in which physical direction it
points.

We therefore introduce a physically informed decision-score framework,
the main methodological contribution of this paper, that translates
probabilistic changepoint detections into signed measures of fault-like
behaviour. For each detected changepoint, the score combines its
posterior probability with its deviation from recent operating
conditions, jump magnitude, local change behaviour, and the physical
direction of the change. In particular, decreases in temperature
difference (the difference between supply and return temperature) and
increases in flow are treated as unfavourable because they are
consistent with reduced heat extraction at the customer installation.
Changes in the opposite direction contribute positively. Utilities
commonly monitor the load factor, the product of flow and temperature
difference, and regard a rise in flow accompanied by a fall in
temperature difference as a sign of insufficient cooling. Applied to the
measurements, however, this product cancels the pattern of interest,
because the increase in flow compensates for the decrease in temperature
difference. We therefore apply the product to the evidence rather than
to the measurements: installations with unfavourable changes in both
channels are ranked first by the product of that evidence, and the
remaining installations by the sum of the channel scores.

The framework is evaluated using operational service records from
Aalborg Forsyning. The evaluation comprises households with isolated
reported fault events together with comparison cases without reported
faults, enabling the ranking produced by the decision score to be
assessed against operational fault information while acknowledging the
limitations of service records as ground truth. The contributions of
this study are therefore threefold: (i) the use of BEAST-based
probabilistic changepoint detection to locate candidate changes in
household heat meter data and to date them without the report date, so
that the dates can be checked against the report rather than derived
from it; (ii) a physically informed decision score that goes beyond the
posterior probability of a change, by weighting it with jump size,
deviation from recent operation and physical direction, so that
installations can be ordered for inspection, and that can equally be
applied to the output of other changepoint detectors; and (iii) an
installation-level ranking framework for fault screening and predictive
maintenance that is evaluated against a large set of operational service
records, on its own and in combination with the level-based screening
that utilities already perform, together with an estimate of the
operational consequences of the detected changes.

\begin{figure}

\centering{

\includegraphics[width=1\linewidth,height=\textheight,keepaspectratio]{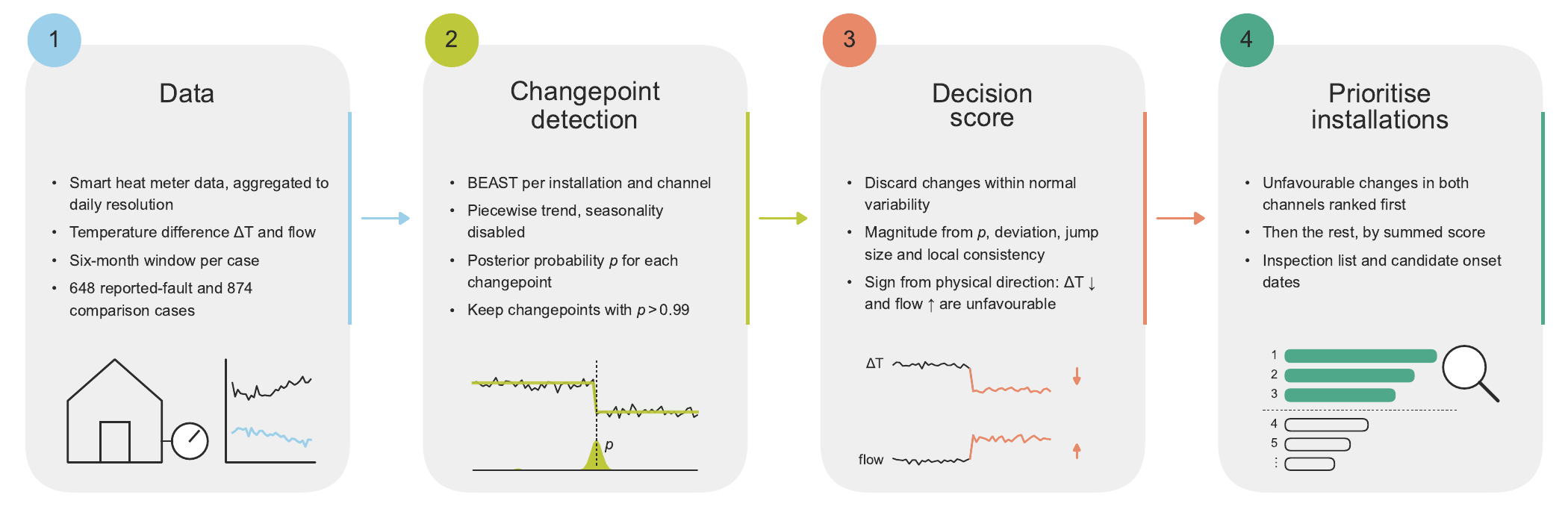}

}

\caption{\label{fig-workflow}Overview of the workflow. (1) Daily heat
meter measurements of \(\Delta T\) and flow are extracted for six-month
observation windows. (2) BEAST detects changepoints in each channel,
retaining those with posterior probability \(p > 0.99\). (3) Each
changepoint receives a signed decision score that combines its posterior
probability with its deviation from recent operation, jump size, local
consistency and physical direction. (4) Installations with unfavourable
changes in both channels are ranked first, followed by the remaining
installations by summed score.}

\end{figure}

Figure~\ref{fig-workflow} summarises the workflow, from the heat meter
data to the ranked list of installations for inspection.

\section{Data}\label{data}

\subsection{Data and cohort
construction}\label{data-and-cohort-construction}

The dataset comprised 807 operational fault reports from 743 households
in the Aalborg district-heating system. Heat meter measurements were
available for 741 of these households. The heat meters record energy,
volume, supply temperature \(T_s\), and return temperature \(T_r\). From
these, we derive the temperature difference \(\Delta T = T_s - T_r\) and
the flow \(\dot V\). Raw measurements were obtained from CenterDenmark
at an hourly resolution. At this resolution, commercial smart heat meter
readings are limited by coarse quantisation, as energy is commonly
recorded in truncated whole kilowatt-hours (18). For the fault cohort,
the measurements covered the period from 3 September 2021 to 10 April
2025.

The recorded date of a fault report is the date of customer contact, not
the date on which the fault began. Because a report can only follow the
appearance of a fault, sometimes by days or weeks, the onset of each
reported fault is unknown and lies at an unknown point before the
recorded date. For each reported fault, a six-month observation window
preceding the recorded contact date was constructed. The window ends at
the report, but it is not a labelled fault period: it may contain a long
stretch of normal operation before the fault begins. Eight households
lacked sufficient historical data and were excluded. The remaining
measurements were aggregated to a 24-hour resolution using the median
when multiple observations were available within the same day, which
also reduces the influence of this quantisation. Data completeness was
then assessed at the daily resolution. Households were retained only if
no more than 5\% of observations were missing over the complete
six-month window and no more than 1\% were missing during the final
month preceding the contact date. These criteria excluded a further 92
households. Missing observations in the retained cases were linearly
interpolated.

Fault events were subsequently thinned to ensure that, for each retained
case, no additional fault had been reported for the same household
during the preceding six-month observation window. This reduced
ambiguity when relating detected structural changes to the reported
fault at the end of the observation period. The resulting fault cohort
comprised 648 fault cases from 641 unique households.

A comparison cohort of 874 cases without reported faults was constructed
using the same observation-window and data-quality criteria. For each
comparison household, an end date was drawn at random, subject to the
requirement that at least six months of preceding measurements were
available. The six months preceding this date were retained as the
observation window. Measurements for these households were, however,
only available from 14 September 2022 to 14 April 2023, so all
comparison windows end between 20 March and 13 April 2023 and cover the
2022/23 heating season, whereas the fault windows end throughout the
year. Results are therefore also reported for the season-matched subset
of cases whose windows end in March or April (965 cases, 91 with a
reported fault). Measurements were aggregated to the same 24-hour
resolution, and missing observations were linearly interpolated using
the same procedure as for the fault cohort.

Comparison cases are defined only by the absence of a service report and
are not verified to be fault-free. Faults that cause no customer
contact, such as gradually deteriorating heat extraction, may remain
unreported, so some comparison cases are expected to contain faults. The
comparison cohort is therefore treated as unlabelled rather than as a
fault-free reference.

\section{Methods}\label{methods}

\subsection{Changepoint detection}\label{changepoint-detection}

Let

\begin{equation}\protect\phantomsection\label{eq-data}{
D_{h,c}=\{(t_i,y_{h,c,i})\}_{i=1}^{n_h}
}\end{equation}

denote the heat meter time series for household \(h\) and sensor channel
\(c\). The analysis is performed separately for each household and
sensor channel. We apply BEAST (17) with the seasonal component
disabled, so that the observed signal is represented by a piecewise
trend and residual noise,

\begin{equation}\protect\phantomsection\label{eq-trend-model}{
y_{h,c,i}=T_{h,c}(t_i;\Theta_{T,h,c})+\varepsilon_{h,c,i}, \qquad \varepsilon_{h,c,i}\sim\mathcal{N}(0,\sigma_{h,c}^2).
}\end{equation}

The trend component is represented by piecewise linear segments, with
structural changes encoded through changes in the underlying trend
structure. Let \(M_{h,c}\) denote the unknown model structure, including
the number and timing of changepoints and the corresponding
segment-specific terms. Conditional on \(M_{h,c}\), the model can be
written as

\begin{equation}\protect\phantomsection\label{eq-linear-model}{
y_{h,c,i}=x_{M_{h,c}}(t_i)^{\intercal}\beta_{M_{h,c}}+\varepsilon_{h,c,i},
}\end{equation}

where \(x_{M_{h,c}}(t_i)\) is the design vector implied by the model
structure and \(\beta_{M_{h,c}}\) contains the corresponding
coefficients. BEAST estimates

\begin{equation}\protect\phantomsection\label{eq-posterior}{
p(M_{h,c},\beta_{M_{h,c}},\sigma_{h,c}^2\mid D_{h,c})\propto p(D_{h,c}\mid M_{h,c},\beta_{M_{h,c}},\sigma_{h,c}^2)p(\beta_{M_{h,c}},\sigma_{h,c}^2\mid M_{h,c})p(M_{h,c}),
}\end{equation}

using Markov chain Monte Carlo with reversible-jump moves that allow the
number and locations of changepoints to vary between posterior samples.
Changepoint inference is subsequently obtained through Bayesian model
averaging.

For each detected changepoint \(\tau_k\), BEAST provides a posterior
probability \(p_{h,c,k}\). Given a probability threshold \(q\), the
candidate changepoints are

\begin{equation}\protect\phantomsection\label{eq-cp-set}{
\widehat{C}_{h,c,q}^{0}=\{\tau_k:p_{h,c,k}>q\}.
}\end{equation}

A threshold of \(q=0.99\) is used for the final fault ranking. Only
changepoints occurring after an initial 30-day reference period are
considered in the subsequent decision-score calculation. Unlike the
penalty of penalised detectors such as PELT (19), which has to be tuned
and has no direct probabilistic meaning, the threshold \(q\) is a
posterior probability and can be set in advance. Appendix A shows that
the results change little for values of \(q\) between 0.5 and 0.999.

\subsection{Physically informed changepoint decision
score}\label{physically-informed-changepoint-decision-score}

A statistically supported changepoint does not necessarily correspond to
an operational fault. We therefore convert each BEAST detection into a
signed decision score that accounts for the magnitude, abnormality,
local behaviour, and physical direction of the observed change.

For a candidate changepoint \(\tau\in\widehat{C}_{h,c,q}^{0}\), define
the first difference

\begin{equation}\protect\phantomsection\label{eq-diff}{
\Delta y_{h,c}(t_i)=y_{h,c,i}-y_{h,c,i-1}.
}\end{equation}

Let \(W_k(t)=\{t-k,\dots,t+k\}\) denote the window of \(\pm k\) days
around \(t\). Because the changepoint location returned by BEAST may not
coincide exactly with the largest observed discontinuity, the location
is refined within a local window of \(\pm1\) day. The adjusted
changepoint is

\begin{equation}\protect\phantomsection\label{eq-refine}{
\widetilde{\tau}=\arg\max_{t\in W_1(\tau)}|\Delta y_{h,c}(t)|.
}\end{equation}

To determine whether the detected change is large relative to the
household's recent operating behaviour, a local reference distribution
is constructed from the 30 days preceding \(\widetilde{\tau}\). Let

\begin{equation}\protect\phantomsection\label{eq-local-mean}{
\mu_{h,c}(\widetilde{\tau})=\operatorname{mean}\{y_{h,c}(t):\widetilde{\tau}-30\text{ days}\leq t<\widetilde{\tau}\},
}\end{equation}

and

\begin{equation}\protect\phantomsection\label{eq-local-sd}{
\sigma_{h,c}(\widetilde{\tau})=\operatorname{sd}\{y_{h,c}(t):\widetilde{\tau}-30\text{ days}\leq t<\widetilde{\tau}\}.
}\end{equation}

The standardised local deviation is defined as

\begin{equation}\protect\phantomsection\label{eq-deviation}{
Z_{h,c}(\widetilde{\tau})=\max_{t\in W_2(\widetilde{\tau})}\frac{|y_{h,c}(t)-\mu_{h,c}(\widetilde{\tau})|}{\max(\sigma_{h,c}(\widetilde{\tau}),\epsilon)},
}\end{equation}

where \(\epsilon>0\) prevents division by zero. Changepoints with

\begin{equation}\protect\phantomsection\label{eq-z-threshold}{
Z_{h,c}(\widetilde{\tau})\leq3
}\end{equation}

are discarded as being within the expected range of normal operating
variability.

For retained changepoints, three quantities determine the magnitude of
the decision score. First, the jump-size term is the size of the
immediate change at the adjusted changepoint,

\begin{equation}\protect\phantomsection\label{eq-jump}{
J_{h,c}(\widetilde{\tau})=|\Delta y_{h,c}(\widetilde{\tau})|,
}\end{equation}

measured in the units of the channel (K for \(\Delta T\) and m³/h for
flow).

Second, the local consistency term compares the net change over the
wider \(\pm2\)-day vicinity with the immediate jump,

\begin{equation}\protect\phantomsection\label{eq-consistency}{
C_{h,c}(\widetilde{\tau})=\frac{\left|\sum_{t\in W_2(\widetilde{\tau})}\Delta y_{h,c}(t)\right|}{\max(|\Delta y_{h,c}(\widetilde{\tau})|,\epsilon)}.
}\end{equation}

The unsigned decision-score magnitude is then

\begin{equation}\protect\phantomsection\label{eq-evidence}{
A_{h,c}(\widetilde{\tau})=p_{h,c,k}\log(1+Z_{h,c}(\widetilde{\tau}))\log(1+J_{h,c}(\widetilde{\tau}))\log(1+C_{h,c}(\widetilde{\tau})),
}\end{equation}

where \(p_{h,c,k}\) denotes the BEAST posterior probability of the
original candidate changepoint \(\tau=\tau_k\).

The score is assigned a sign according to the physical direction of the
change. The final decision score uses the temperature difference
\(\Delta T\) and the flow. For \(\Delta T\), a decrease is considered
unfavourable,

\begin{equation}\protect\phantomsection\label{eq-sign-dt}{
s_{h,\Delta T}(\widetilde{\tau})= \begin{cases}
-1, & \Delta y_{h,\Delta T}(\widetilde{\tau}) < 0, \\
+1, & \Delta y_{h,\Delta T}(\widetilde{\tau}) \geq 0,
\end{cases}
}\end{equation}

whereas for flow, an increase is considered unfavourable,

\begin{equation}\protect\phantomsection\label{eq-sign-flow}{
s_{h,\mathrm{flow}}(\widetilde{\tau})= \begin{cases}
-1, & \Delta y_{h,\mathrm{flow}}(\widetilde{\tau}) > 0, \\
+1, & \Delta y_{h,\mathrm{flow}}(\widetilde{\tau}) \leq 0.
\end{cases}
}\end{equation}

The signed contribution from changepoint \(\widetilde{\tau}\) is
therefore

\begin{equation}\protect\phantomsection\label{eq-signed-score}{
d_{h,c}(\widetilde{\tau})=s_{h,c}(\widetilde{\tau})A_{h,c}(\widetilde{\tau}).
}\end{equation}

Return temperature is not scored separately: at a given supply
temperature, a rise in return temperature appears as a fall in
\(\Delta T\), so the two channels largely carry the same information.
Flow, in contrast, provides a second, physically distinct symptom of
insufficient cooling. Energy, volume, and supply and return temperatures
were retained in diagnostic plots to provide context for the overall
behaviour of each household.

\subsection{Installation-level ranking}\label{sec-ranking}

The choice of how to combine the two channels is motivated by the load
factor \(\dot V\,\Delta T\), the product of flow \(\dot V\) and
\(\Delta T\), which is proportional to the delivered heat and is
monitored by utilities. A rise in flow accompanied by a fall in
\(\Delta T\) indicates that the installation draws more water without
cooling it sufficiently, which raises return temperatures in the
network. In the load factor itself, however, the two changes largely
offset each other, because the building's heat demand is unchanged by
the fault. The pattern of interest is therefore the joint occurrence of
unfavourable changes in both channels rather than a change in their
product.

For each household and channel, let

\begin{equation}\protect\phantomsection\label{eq-channel-sums}{
S_{h,c}=\sum_{\widetilde{\tau}}d_{h,c}(\widetilde{\tau}), \qquad N_{h,c}=\sum_{\widetilde{\tau}}\min\{d_{h,c}(\widetilde{\tau}),0\}
}\end{equation}

denote the total signed score and the accumulated unfavourable evidence,
respectively. Two installation-level quantities are formed: the joint
evidence

\begin{equation}\protect\phantomsection\label{eq-joint-evidence}{
P_h=|N_{h,\Delta T}|\,|N_{h,\mathrm{flow}}|,
}\end{equation}

and the summed score

\begin{equation}\protect\phantomsection\label{eq-summed-score}{
S_h=S_{h,\Delta T}+S_{h,\mathrm{flow}}.
}\end{equation}

Cases with \(P_h>0\), that is, with unfavourable changes in both
\(\Delta T\) and flow, form the first priority tier and are ranked in
descending order of \(P_h\). The remaining cases, with unfavourable
evidence in at most one channel, follow in ascending order of \(S_h\).
Positions nearer the top of the resulting ranking therefore indicate
stronger evidence of fault-like behaviour, with joint evidence in both
channels taking precedence over evidence in a single channel.

\section{Results}\label{results}

\subsection{Fault prioritisation
performance}\label{fault-prioritisation-performance}

All evaluation cases were ranked as described in
Section~\ref{sec-ranking}, with joint evidence in \(\Delta T\) and flow
first and the remaining cases by their summed score. Of the 1,522 cases,
407 showed unfavourable changes in both channels and formed the first
priority tier; 331 (81\%) of these were associated with a reported
fault. Reported faults were highly concentrated among the
highest-priority cases (Table~\ref{tbl-fault-ranking}).

\begin{longtable}[]{@{}ccc@{}}
\caption{Fault prioritisation performance of the decision score. The
table reports the number of cases associated with a reported fault among
the \(K\) highest-priority cases and their proportion, Precision@K.
High-priority cases with no reported fault are counted as non-faults,
although some may contain unreported
faults.}\label{tbl-fault-ranking}\tabularnewline
\toprule\noalign{}
Highest-priority cases & Cases with reported fault & Precision@K \\
\midrule\noalign{}
\endfirsthead
\toprule\noalign{}
Highest-priority cases & Cases with reported fault & Precision@K \\
\midrule\noalign{}
\endhead
\bottomrule\noalign{}
\endlastfoot
25 & 24 & 0.96 \\
50 & 47 & 0.94 \\
100 & 94 & 0.94 \\
200 & 181 & 0.91 \\
300 & 261 & 0.87 \\
\end{longtable}

The share of reported faults stayed at 0.87 or higher up to the 300
highest-priority cases.

\begin{figure}

\centering{

\includegraphics[width=0.65\linewidth,height=\textheight,keepaspectratio]{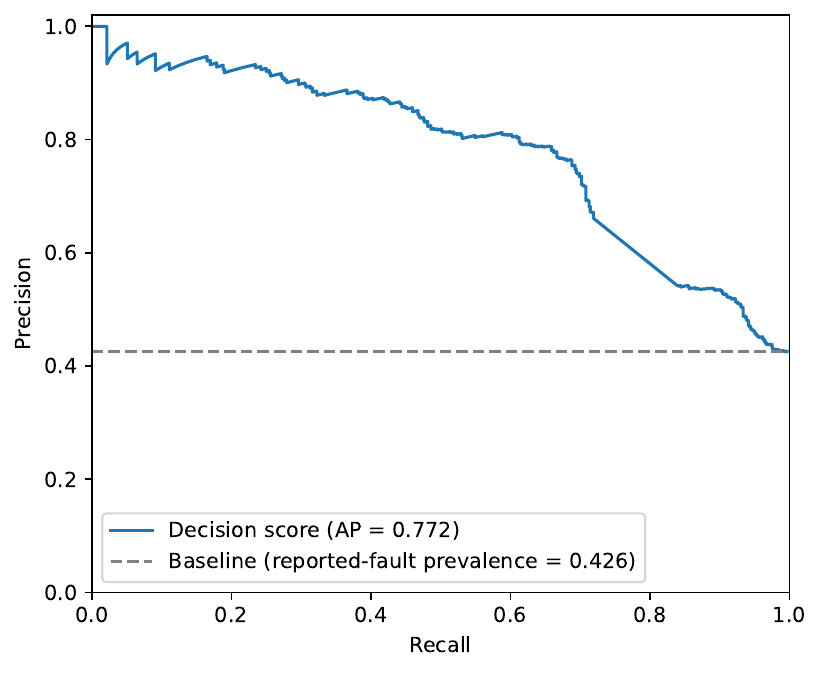}

}

\caption{\label{fig-pr-curve}Precision--Recall curve for fault
prioritisation across all 1,522 evaluation cases. The decision score
achieved an AP of 0.772, compared with a baseline reported-fault
prevalence of 0.426.}

\end{figure}

Overall ranking performance was evaluated across all 1,522 cases using
the Precision--Recall curve (Figure~\ref{fig-pr-curve}). Of these, 648
cases (42.6\%) were associated with a reported fault. The curve is
summarised by the Average Precision (AP), the mean of the precision
values at the ranks of the reported-fault cases. The decision score
achieved an AP of 0.772, compared with the baseline reported-fault
prevalence of 0.426. Together, these results show that the decision
score effectively concentrates reported faults among the
highest-priority cases.

Both Precision@K and AP are computed against service reports. A
comparison case that receives a low score is counted as a non-fault,
even though it may contain a fault that was never reported. These
metrics therefore measure agreement with the service records and are
likely conservative estimates of how well the ranking prioritises actual
faults.

\subsection{Timing of detected changes}\label{sec-timing}

The ranking does not use the position of a changepoint within the
observation window, so the timing of the detected changes provides an
independent check on whether they relate to the reported faults. For
every case with unfavourable evidence in a channel, we located the
changepoint with the most negative contribution and measured its
distance to the end of the window. For reported-fault cases, the
strongest unfavourable change in flow occurred a median of 11 days
before the report, and 70\% of these changes fell within the final 30
days. For \(\Delta T\), the median was 26 days and 52\% fell within the
final 30 days. If the changes were spread uniformly over the scored part
of the window, about 20\% would fall within the final 30 days.
Comparison cases, whose windows end at an arbitrary date, showed no such
concentration: the medians were 128 days for flow and 87 days for
\(\Delta T\), with 4\% and 18\% of the changes within the final 30 days.

The detected changes therefore cluster in the weeks before the report,
where the onset of a reported fault is expected to lie. This clustering
is not built into the method: neither the detection nor the ranking uses
the report date or the position of a changepoint within the window, and
the same procedure finds no such clustering in the comparison cases,
whose windows end at an arbitrary date. The dates are recovered from the
measurements alone.

\subsection{Combination with level-based
screening}\label{sec-combination}

Utilities already screen for poor cooling by ranking installations by
the level of \(\Delta T\) (1,20). Ranked by the mean \(\Delta T\) over
the final 30 days of the window, the level separates the groups well
(area under the receiver operating characteristic (ROC) curve, AUC,
0.909, against 0.798 for the decision score; Table~\ref{tbl-level}), but
it cannot tell when the low \(\Delta T\) began. The decision score
carries complementary information: that an installation's operation
changed, and when. Summing the ranks of the two rankings, without any
weighting or tuning, ranked the cases better than either alone. On all
cases, AUC rose to 0.929 and AP to 0.905. In the season-matched subset,
where reported faults make up 9.4\% of the cases, AP rose from 0.486 for
level screening and 0.446 for the decision score to 0.646, a gain over
level screening of 0.160 (95\% interval 0.060 to 0.257). On all cases,
the gain was 0.033 (0.011 to 0.054). Both intervals come from a paired
bootstrap over cases with 2,000 resamples. In the season-matched subset,
the share of reported faults among the 25 highest-priority cases also
rose, from 0.60 for level screening to 0.80, and the gain held in 74 of
the 75 parameter settings and BEAST seeds examined in Appendix A.

For a utility, this means fewer inspections. In the season-matched
subset, inspecting the 59 installations ranked highest by the
combination (6\% of the 965) would find half of the 91 reported faults.
Level screening alone would need 77 inspections (8\%) to find the same
number, and a random order about 480 (50\%). To find 90\% of the
reported faults, the combination would need 218 inspections (23\%),
against 355 (37\%) for level screening alone.

\begin{longtable}[]{@{}
  >{\raggedright\arraybackslash}p{(\linewidth - 12\tabcolsep) * \real{0.3465}}
  >{\centering\arraybackslash}p{(\linewidth - 12\tabcolsep) * \real{0.1089}}
  >{\centering\arraybackslash}p{(\linewidth - 12\tabcolsep) * \real{0.1089}}
  >{\centering\arraybackslash}p{(\linewidth - 12\tabcolsep) * \real{0.1089}}
  >{\centering\arraybackslash}p{(\linewidth - 12\tabcolsep) * \real{0.1089}}
  >{\centering\arraybackslash}p{(\linewidth - 12\tabcolsep) * \real{0.1089}}
  >{\centering\arraybackslash}p{(\linewidth - 12\tabcolsep) * \real{0.1089}}@{}}
\caption{Decision score compared with level-based \(\Delta T\)
screening, for all cases and for the season-matched cases whose windows
end in March or April (965 cases, 91 with a reported fault). The level
rankings order the cases by their mean \(\Delta T\) over the final or
the first 30 days of the window, lowest first. The combined ranking sums
the ranks of the decision score and of the final-30-day level, without
weighting or tuning.}\label{tbl-level}\tabularnewline
\toprule\noalign{}
\begin{minipage}[b]{\linewidth}\raggedright
Ranking
\end{minipage} & \begin{minipage}[b]{\linewidth}\centering
All: AUC
\end{minipage} & \begin{minipage}[b]{\linewidth}\centering
All: AP
\end{minipage} & \begin{minipage}[b]{\linewidth}\centering
Matched: AUC
\end{minipage} & \begin{minipage}[b]{\linewidth}\centering
Matched: AP
\end{minipage} & \begin{minipage}[b]{\linewidth}\centering
Matched: P@25
\end{minipage} & \begin{minipage}[b]{\linewidth}\centering
Matched: P@50
\end{minipage} \\
\midrule\noalign{}
\endfirsthead
\toprule\noalign{}
\begin{minipage}[b]{\linewidth}\raggedright
Ranking
\end{minipage} & \begin{minipage}[b]{\linewidth}\centering
All: AUC
\end{minipage} & \begin{minipage}[b]{\linewidth}\centering
All: AP
\end{minipage} & \begin{minipage}[b]{\linewidth}\centering
Matched: AUC
\end{minipage} & \begin{minipage}[b]{\linewidth}\centering
Matched: AP
\end{minipage} & \begin{minipage}[b]{\linewidth}\centering
Matched: P@25
\end{minipage} & \begin{minipage}[b]{\linewidth}\centering
Matched: P@50
\end{minipage} \\
\midrule\noalign{}
\endhead
\bottomrule\noalign{}
\endlastfoot
Decision score & 0.798 & 0.772 & 0.855 & 0.446 & 0.64 & 0.56 \\
\(\Delta T\) level, final 30 days & 0.909 & 0.872 & 0.901 & 0.486 & 0.60
& 0.60 \\
\(\Delta T\) level, first 30 days & 0.604 & 0.546 & 0.649 & 0.148 & 0.16
& 0.10 \\
Combined (sum of ranks) & 0.929 & 0.905 & 0.945 & 0.646 & 0.80 & 0.82 \\
\end{longtable}

The evaluation favours both rankings that use the level, because each
fault window ends at the report, when the fault is fully developed. The
gain over level screening alone, rather than the combined figures
themselves, is therefore the relevant measure. In practice, the decision
score is best used alongside the level screening that utilities already
perform, adding to it the evidence that, and when, an installation's
operation changed.

\subsection{Operational consequences of the detected
changes}\label{sec-consequences}

The detected changes also have a measurable operational cost. For the
reported-fault cases in the first priority tier, we compared the 30 days
before the strongest unfavourable \(\Delta T\) change with the period
from the change to the report, using the 254 of these 331 cases with at
least seven days after the change (median 38 days). Over that period,
the median installation's \(\Delta T\) fell by 15 K and its return
temperature rose by 16 K, and it circulated twice as much water per day
as before. Delivering the same heat at the installation's earlier
\(\Delta T\) would have needed about half of that water, so the network
circulated a median of 1.2 m³ per installation and day that delivered no
heat. Comparison cases in the first tier, measured in the same way,
showed a fall in \(\Delta T\) of only 3 K, and only 9\% of the water
they circulated after the change delivered no heat, so season and
weather are unlikely to explain the effect. Among the 100
highest-priority installations of the combined ranking in the
season-matched subset, 82 had such a change, and for the median
installation 40\% of the water circulated after it delivered no heat.
These changes carry an energy cost: a 16 K rise in return temperature
increases heat losses from the return pipes and lowers the efficiency of
heat production, and doubling the circulated flow raises the pumping
energy needed to deliver the same heat (1,2).

\subsection{Robustness to modelling choices}\label{sec-robustness}

The decision score rests on choices that are not estimated from the
data: the BEAST probability threshold \(q\), the
three-standard-deviation criterion, the 30-day reference period, the
widths of the local windows, the four factors of \(A_{h,c}\), the rule
for combining \(\Delta T\) and flow, and the random seed of the BEAST
sampler. Appendix A varies each of these and repeats the evaluation, and
none of them changes the results qualitatively. Over all parameter
settings examined, one at a time and in combination, AP ranged from 0.67
to 0.79 against a reported-fault prevalence of 0.426, at least 92\% of
the 25 highest-priority cases had a reported fault, and the strongest
unfavourable flow change in reported-fault cases fell a median of 9 to
13 days before the report. Changing a single parameter within a moderate
range (\(q\) from 0.5 to 0.999, a criterion between 2.5 and 4 standard
deviations, reference periods of 21 to 60 days, a local window \(W_2\)
of \(\pm1\) to \(\pm5\) days, and a refinement window of \(\pm1\) or
\(\pm2\) days) moved AP by at most 0.034 from the value of 0.772
reported above. Several of these settings performed slightly better than
the one used here, so the results do not depend on tuning the parameters
to the evaluation data. Re-running BEAST with three other random seeds
changed AP by at most 0.005 and left the median timing of the flow
changes at 11 days.

The ablations in Appendix A show which elements carry the ranking. The
largest losses came from dropping the three-standard-deviation criterion
(AP 0.702) or the jump-size term (0.728), and the two-tier rule achieved
a higher AP than ranking by \(\Delta T\) or flow alone, by their plain
sum, or by the joint evidence alone. Ranking by the signed BEAST
posterior probabilities alone gave an AP of 0.542, so the ranking is
driven by the physically informed post-processing rather than by the
strength of the statistical evidence for a change. Two elements matter
less in this dataset. The posterior probability acts in practice as a
threshold rather than as a weight, because the probabilities of the
retained changepoints are close to one, and the local consistency term
\(C\) leaves AP unchanged but raises the share of reported faults among
the 100 highest-priority cases from 0.91 to 0.94.

The decision score is not tied to BEAST. Appendix A applies the same
score and ranking rule to the changepoints of PELT (19), a widely used
penalised detector, and to those of BEAST with a piecewise-constant
rather than a piecewise-linear trend. With every detector and setting,
reported faults remained concentrated among the highest-priority cases,
with at least 90\% of the 100 highest-priority cases associated with a
reported fault. Detectors that model shifts in the level, the kind of
change that the decision score measures, ranked the cases somewhat
better than the piecewise-linear trend used here: BEAST with a constant
trend reached an AP of 0.790 at the same threshold, and PELT 0.733 to
0.807 depending on its penalty. BEAST, however, dated the changes more
sharply. With either trend model and the threshold used here, the
strongest unfavourable flow change in reported-fault cases fell a median
of 11 days before the report, against 12 to 17 days for PELT. Among
detectors that model linear trends, BEAST ranked the cases considerably
better than PELT at a comparable number of changepoints (AP 0.772
against 0.638 to 0.679).

\subsection{Household-level examples}\label{household-level-examples}

To illustrate how the decision score relates to the underlying heat
meter measurements, three cases were selected from different positions
within the 300 highest-priority cases. The diagnostic plots show energy,
volume, supply temperature, return temperature, temperature difference,
and flow, together with the decision-score contributions of the detected
changepoints. Only temperature difference and flow contribute to the
ranking. All three cases belong to the first priority tier.

\begin{figure}[H]

\centering{

\includegraphics[width=0.8\linewidth,height=\textheight,keepaspectratio]{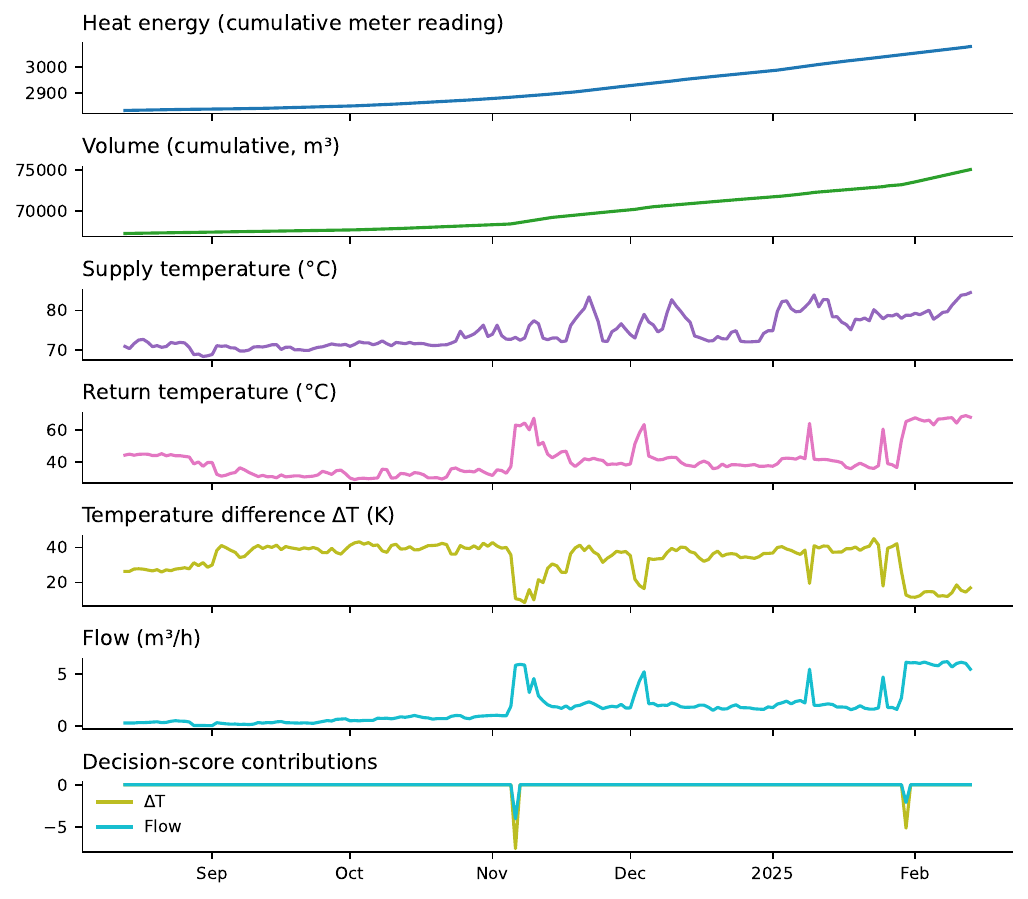}

}

\caption{\label{fig-1}Heat meter measurements for the highest-priority
case (rank 1; joint evidence \(P_h = 78.26\)).}

\end{figure}

The highest-priority case, shown in Figure~\ref{fig-1}, has joint
evidence \(P_h = 78.26\) and is associated with a reported fault. On two
occasions, \(\Delta T\) falls from approximately 40 K to 10 K while the
flow rises several-fold, and the second episode persists until the end
of the observation window.

\begin{figure}[H]

\centering{

\includegraphics[width=0.8\linewidth,height=\textheight,keepaspectratio]{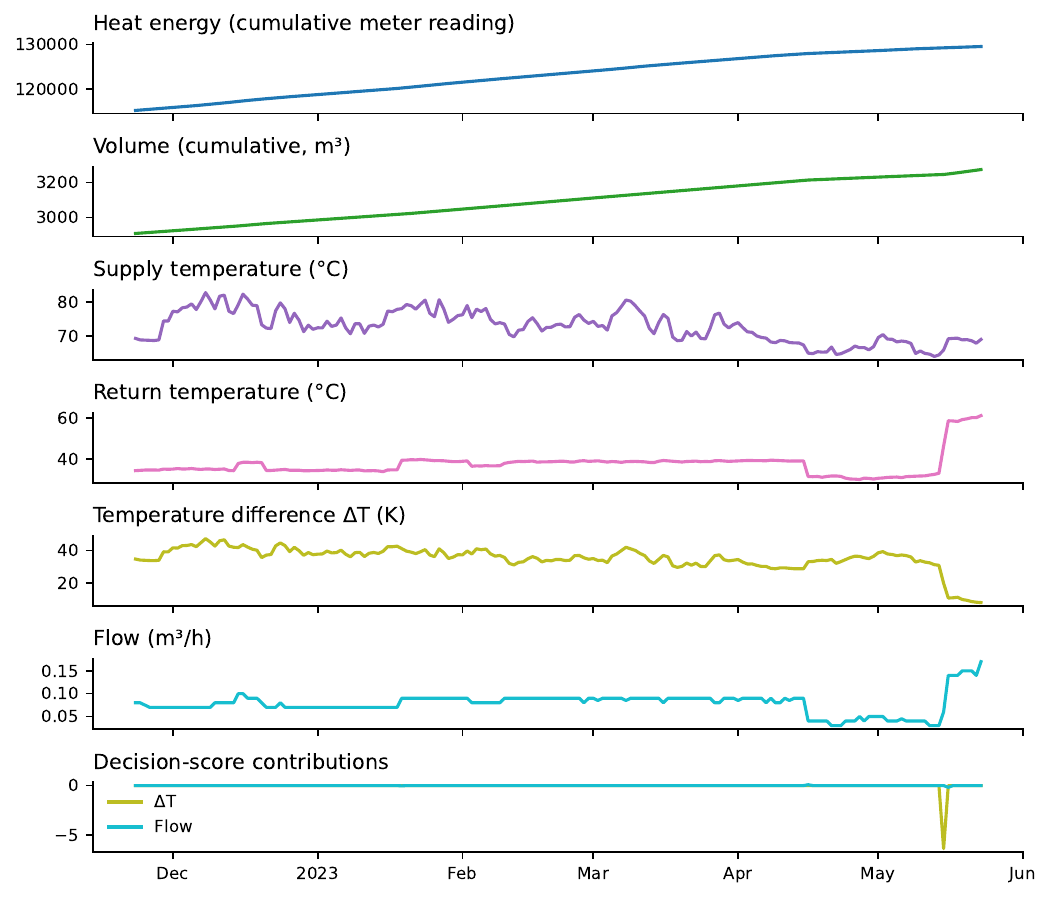}

}

\caption{\label{fig-153}Heat meter measurements for the case ranked
153rd (joint evidence \(P_h = 1.34\)).}

\end{figure}

An intermediate-ranked case is shown in Figure~\ref{fig-153}. This case
is ranked 153rd, has joint evidence \(P_h = 1.34\), and is associated
with a reported fault. Shortly before the end of the observation window,
the flow increases and \(\Delta T\) decreases simultaneously,
accompanied by a rise in return temperature.

\begin{figure}[H]

\centering{

\includegraphics[width=0.8\linewidth,height=\textheight,keepaspectratio]{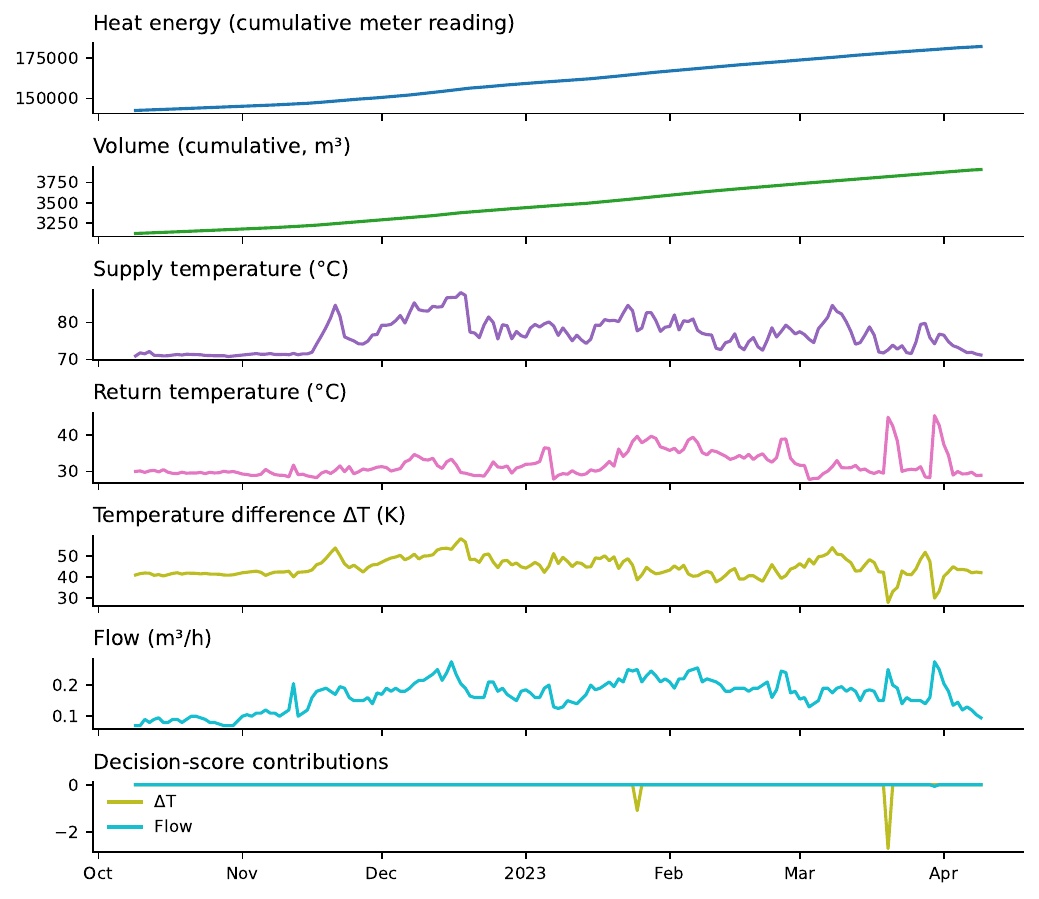}

}

\caption{\label{fig-300}Heat meter measurements for the case ranked
300th, a comparison case without a reported fault (joint evidence
\(P_h = 0.27\)).}

\end{figure}

The case shown in Figure~\ref{fig-300} is ranked 300th and has joint
evidence \(P_h = 0.27\). It is a comparison case without a reported
fault. Nevertheless, it shows two short episodes in which the flow rises
and \(\Delta T\) falls while the return temperature increases, the same
pattern as in the reported-fault cases. Such cases illustrate why
high-priority comparison cases should not automatically be regarded as
false positives.

Together, the examples illustrate the heat meter behaviour associated
with different positions in the ranking, from the highest-priority case
to the boundary of the 300 highest-priority cases.

\section{Discussion}\label{discussion}

The proposed decision-score framework concentrated reported faults among
the highest-priority cases (Table~\ref{tbl-fault-ranking}), which
supports its use for selecting a manageable subset of installations for
inspection rather than requiring a binary classification of every case
(11--14).

Level screening of \(\Delta T\) ranked the reported faults better than
the decision score alone (Section~\ref{sec-combination}), but the design
of the evaluation favours it. Each fault window ends at the report, when
the fault is fully developed, so the evaluation reads the level exactly
where and when a fault is known to be present. The low \(\Delta T\) is,
moreover, largely recent: five to six months before the report, the same
level separates the two groups only weakly (AUC 0.604;
Table~\ref{tbl-level}). Level screening therefore detects the same
recent deterioration as the changepoints, but it cannot tell when the
deterioration began.

This points to what the proposed method adds. The decision score finds
the change without using the report date, places it in the weeks before
the report, and attaches to it a date, a direction and a posterior
probability. These are the ingredients needed to date a fault and to
label data for classification. A level threshold also cannot distinguish
an installation whose \(\Delta T\) has just fallen from one that has
always operated at that level, whereas the decision score measures each
change against the installation's own recent operation. Finally, the two
kinds of evidence are complementary rather than competing, and their
combination ranked the cases best (Section~\ref{sec-combination}).

A central aspect of the framework is the distinction between statistical
changepoints and operational faults. BEAST identifies structural changes
based on posterior evidence for changes in the underlying trend
structure (17), but such changes may also arise from variations in
demand, control behaviour, outdoor conditions, or normal system
operation (1,5,6). The decision score therefore acts as a
post-processing layer that evaluates statistically supported
changepoints according to their magnitude, deviation from recent
operating conditions, local behaviour, and physical direction. This
layer is not tied to BEAST: applied to the changepoints of detectors
that model shifts in the level, the kind of change the score measures,
it ranked the cases about as well or slightly better
(Section~\ref{sec-robustness}). BEAST's advantages lie in a threshold
that is a posterior probability rather than a tuned penalty, and in a
sharper dating of the detected changes. A piecewise-constant trend model
in BEAST combines both and is a natural refinement of the framework.
More generally, because the score only requires the locations of
candidate changes, alternative detectors can be added to the framework,
either in place of BEAST or alongside it, for example to pool the
candidates of detectors that respond to different kinds of change.

The direction of change is particularly important for distinguishing
potentially problematic behaviour from other structural changes.
Decreasing temperature difference and increasing flow are associated
with poorer heat extraction and therefore contribute negatively to the
score. At a given supply temperature, a decreasing temperature
difference corresponds to an increasing return temperature, consistent
with previous analyses of operational fault data that identified return
temperature as a key indicator (15). The use of physically interpretable
variables and domain knowledge has more generally been emphasised in
district-heating fault-detection studies (5,7,8,10,15).

The two channels are not equally informative on their own. A decrease in
\(\Delta T\) at unchanged flow already implies that less heat is
extracted from the supplied water, whereas an increase in flow at
unchanged \(\Delta T\) more often reflects higher heat demand, for
example during cold periods or high domestic hot-water use (21). The
strongest evidence arises when both occur together, which is the pattern
that utilities associate with insufficient cooling. The two-tier ranking
reflects this ordering: joint evidence takes precedence, and among the
remaining cases the summed score allows a decrease in \(\Delta T\) to
count even without a corresponding change in flow. Ranking by the load
factor itself would not capture this pattern, because a rise in flow and
a fall in \(\Delta T\) offset each other in the product of the
measurements.

The score should be interpreted as a relative prioritisation measure
rather than as a calibrated probability of fault. A high joint evidence
\(P_h\) or a highly negative summed score may result from one pronounced
structural change or from several unfavourable contributions accumulated
during the observation period. Consequently, joint evidence of \(100\)
does not imply a fixed increase in fault probability relative to joint
evidence of \(10\); it indicates stronger accumulated evidence of
fault-like behaviour. This distinction is important because fault
severity, statistical confidence, and operational relevance are not
necessarily represented by the same quantity (6,13,14).

The timing of faults is a further reason to treat reported faults with
care. Because a report is created after the fault has appeared, the
recorded date marks the end of an unknown period of faulty operation,
and the observation window may contain a substantial stretch of normal
operation before the onset. Labelling the whole window, or the period
around the report date, as faulty would therefore contaminate a fault
class with non-faulty data, while treating the period before the report
as normal would place early faulty operation in the healthy class. Both
are a direct obstacle to training classifiers that distinguish fault
types. The detected changepoints suggest a way forward, since they
locate abrupt changes in time independently of the report date and the
physical direction of the change indicates whether the change is
fault-like. The report date was not used to find them, yet the strongest
unfavourable changes in the reported-fault cases clustered in the weeks
before the report, whereas those in the comparison cases did not. The
present evaluation does not, however, assess how accurately the detected
changepoints date the onset of a fault, because no ground-truth onset
dates exist. This would require validation, for example against
inspected cases or synthetic faults with known onset (16), before the
changepoints can be used to label data for classification.

The operational fault records also limit the interpretation of the
evaluation. Service reports constitute imperfect ground truth: faults
may be reported after the underlying problem emerges, some abnormal
conditions may never be reported, and recorded fault categories may not
correspond directly to visible changes in the selected sensor variables
(9,14,15,22). A high-priority case without a reported fault should
therefore not automatically be regarded as a false positive (see
Figure~\ref{fig-300}), while a reported fault may not necessarily
produce a clear structural change in temperature difference or flow. In
particular, the comparison cohort is unlabelled rather than fault-free.
Precision@K and AP thus measure agreement with service records rather
than detection accuracy. If unreported faults among the comparison cases
resemble reported ones, these metrics understate performance but
preserve the relative ordering of alternative scoring formulations. A
formulation that preferentially identified faults that are rarely
reported would, however, be penalised by this evaluation.

The decision score itself is physically motivated but remains a
constructed scoring rule. Its components and thresholds, including the
BEAST probability threshold, the three-standard-deviation criterion, the
local time windows, and the rule for combining \(\Delta T\) and flow,
are not estimated from an explicit probabilistic fault model. The
sensitivity analysis and ablations in Appendix A show that the
conclusions do not hinge on these choices, and that the
three-standard-deviation criterion, the jump size and the two-tier rule
carry most of the ranking performance (Section~\ref{sec-robustness}).
These checks were, however, made on the same data on which the score was
developed, and validation on independent data is needed to establish
that the formulation transfers to other district-heating systems.

The analysis is further based on daily aggregated and linearly
interpolated measurements from a single district-heating system.
Aggregation to a 24-hour resolution may facilitate detection of
persistent structural changes by suppressing short-term variability, but
may also obscure faults occurring over shorter time scales. In addition,
BEAST was applied without an explicit seasonal component, so some
demand- or season-related behaviour, which is pronounced in smart heat
meter data (3), may still be represented as structural change. All
comparison windows cover the 2022/23 heating season, whereas the fault
windows end throughout the year, which may leave differences in seasonal
or operating conditions between the two groups. In the season-matched
subset of windows ending in March or April (Table~\ref{tbl-level}), the
decision score separated the groups better than in the full evaluation
(AUC 0.855 against 0.798), so the seasonal imbalance does not appear to
inflate the results. A comparison cohort with end dates matched to the
report dates would nevertheless be preferable. Sensitivity to data
quality, temporal resolution, and operating context has likewise been
highlighted in previous district-heating studies (5,9,16).

\section{Conclusion}\label{conclusion}

Combining BEAST changepoint detection with physically informed
post-processing produced an interpretable ranking of district-heating
cases according to fault-like behaviour. Prioritising installations with
joint unfavourable changes in temperature difference and flow, the
pattern that utilities associate with insufficient cooling, concentrated
reported faults strongly among the highest-priority cases, and the
framework achieved an Average Precision of 0.772 compared with a
baseline reported-fault prevalence of 0.426. Because cases without a
service report are not necessarily fault-free, this agreement with
service records is likely a conservative measure of performance. The
flagged changes also matter operationally: in the reported-fault
installations of the first priority tier, about half of the water
circulated after the detected change delivered no heat.

Level-based \(\Delta T\) screening, as utilities already practise it,
ranks the cases well, partly because the evaluation tells it where to
look, but it cannot tell when a fault began. The proposed framework
locates the change without the report date: in reported-fault cases, the
strongest unfavourable flow change fell a median of 11 days before the
report, against 128 days before the window end in comparison cases. The
two kinds of evidence are complementary: combined, they ranked the cases
better than either alone, which makes the decision score most useful as
an addition to the level screening that utilities already perform.

The framework is intended as a screening and prioritisation tool rather
than a direct fault-probability model. The algorithm is designed to
identify periods of abnormal operation that may warrant further
inspection, and to provide a manageable subset of installations for
operational follow-up. Because poorly cooling installations raise return
temperatures and circulate water that delivers no heat, finding them
sooner shortens the period in which they waste energy in the network.
The decision score is interpretable in terms of the underlying heat
meter measurements and can be used to visualise the evidence for each
detected changepoint. Because reports follow the onset of a fault by an
unknown delay, correct dating of the underlying fault is a necessary
prerequisite for training and validating fault classifiers without
contaminating fault classes with normal operation or the healthy class
with early faulty operation, and the proposed framework can serve as a
first step in such a workflow. Since the decision score is not tied to
BEAST, the framework can also incorporate alternative changepoint
detectors. Further work should validate the dating of fault onset from
the detected changepoints, assess the robustness of the scoring
formulation across temporal resolutions and independent district-heating
datasets, examine which detectors, or combinations of detectors, provide
the best candidate changes, and evaluate the framework for prospective
fault prioritisation in operational settings.

\subsection*{Funding}\label{funding}
\addcontentsline{toc}{subsection}{Funding}

The D4Heat project is funded by the Energy Technology Development and
Demonstration Programme (EUDP; Case no. 640241-521759).

\subsection*{Author contributions}\label{author-contributions}
\addcontentsline{toc}{subsection}{Author contributions}

T.K.B.: Methodology, Writing -- original draft, Writing -- review \&
editing. N.P.B.P.: Conceptualization, Methodology, Software, Formal
analysis. J.E.V.-V.: Conceptualization, Methodology, Software, Formal
analysis, Supervision, Writing -- review \& editing. M.Z.P.:
Conceptualization, Supervision, Writing -- review \& editing. A.M.-P.:
Resources, Project administration, Funding acquisition.

\subsection*{Data availability}\label{data-availability}
\addcontentsline{toc}{subsection}{Data availability}

The household-level heat meter data and service records used in this
study are subject to a confidentiality agreement within the D4Heat
project and cannot be shared publicly. Aggregated results and derived
summaries are reported in the paper. Synthetic heat meter data that
reproduce the workflow end to end are archived on Zenodo with the code
(23).

\subsection*{Code availability}\label{code-availability}
\addcontentsline{toc}{subsection}{Code availability}

Changepoint detection was performed with the Rbeast package (24). An
installable Python implementation of the decision score and ranking
rule, which accepts the changepoints of any detector, is available at
\url{https://github.com/everval/dh-fault-score} and archived on Zenodo
(23), together with synthetic heat meter data that reproduce the
workflow end to end, since the data of this study cannot be shared. Its
scores are identical to those of the analysis code used for this paper.

\subsection*{Declaration of competing
interest}\label{declaration-of-competing-interest}
\addcontentsline{toc}{subsection}{Declaration of competing interest}

The authors declare that they have no known competing financial
interests or personal relationships that could have appeared to
influence the work reported in this paper.

\subsection*{Declaration of generative AI and AI-assisted technologies
in the writing
process}\label{declaration-of-generative-ai-and-ai-assisted-technologies-in-the-writing-process}
\addcontentsline{toc}{subsection}{Declaration of generative AI and
AI-assisted technologies in the writing process}

During the preparation of this work the authors used Claude (Anthropic)
to assist with proofreading, and with the preparation of the
accompanying function repository. After using this tool, the authors
reviewed and edited the content as needed and take full responsibility
for the content of the publication.

\subsection*{Ethics and privacy}\label{ethics-and-privacy}
\addcontentsline{toc}{subsection}{Ethics and privacy}

The analysis was conducted on anonymised household-level data. No
directly identifying household information is reported.

\section*{References}\label{references}
\addcontentsline{toc}{section}{References}

\protect\phantomsection\label{refs}

\newpage{}

\section*{Appendix A. Sensitivity and ablation of the decision
score}\label{appendix-a.-sensitivity-and-ablation-of-the-decision-score}
\addcontentsline{toc}{section}{Appendix A. Sensitivity and ablation of
the decision score}

The decision score and the ranking rule involve choices that were fixed
in advance rather than estimated from the data. This appendix examines
whether these choices drive the results. The question is not whether
another setting would score slightly higher, but whether any reasonable
setting changes the paper's four conclusions. These are that reported
faults are concentrated among the highest-priority cases, that AP is
well above the reported-fault prevalence, that the results are not
driven by the seasonal imbalance between the cohorts, and that the
strongest unfavourable flow changes in reported-fault cases fall in the
weeks before the report.

\subsection*{Set-up}\label{set-up}
\addcontentsline{toc}{subsection}{Set-up}

Every variation repeats the evaluation of the Results section. The
parameter, component and ranking-rule variations re-score the
changepoints of the main analysis; the seed and detector checks run the
changepoint detection again.

\emph{Parameters.} Each parameter was varied with the others held at the
values of the main analysis (in parentheses):

\begin{itemize}
\tightlist
\item
  the BEAST probability threshold
  \(q\in\{0.5,0.7,0.8,0.9,0.95,0.99,0.999\}\) (0.99);
\item
  the criterion \(Z_{h,c}(\widetilde{\tau})>z\), with
  \(z\in\{1,2,2.5,3,3.5,4,5\}\) or no criterion (3);
\item
  the reference period \(L\in\{14,21,30,45,60\}\) days, used both as the
  initial period in which changepoints are not scored and as the window
  before \(\widetilde{\tau}\) from which \(\mu_{h,c}\) and
  \(\sigma_{h,c}\) are computed (30 days);
\item
  the window before \(\widetilde{\tau}\) alone, \(\{7,14,21,30\}\) days,
  with the initial period kept at 30 days;
\item
  the half-width of the local window \(W_2\) used for \(Z\) and \(C\),
  \(\{1,2,3,5\}\) days (2 days);
\item
  the half-width of the refinement window, \(\{0,1,2,3\}\) days, where 0
  means no refinement (1 day).
\end{itemize}

To check for interactions, \(q\in\{0.9,0.95,0.99,0.999\}\),
\(z\in\{2,3,4\}\) and \(L\in\{21,30,45\}\) were also varied jointly,
giving 36 combinations.

\emph{Score components.} Each factor of \(A_{h,c}\), namely \(p\),
\(\log(1+Z)\), \(\log(1+J)\) and \(\log(1+C)\), was replaced by 1 in
turn. The criterion \(Z>3\) was removed, and the cases were ranked by
the posterior probability alone, signed by the direction of the change,
with and without the criterion.

\emph{Ranking rule.} The two-tier rule was compared with ranking by
\(S_{h,\Delta T}\) alone, by \(S_{h,\mathrm{flow}}\) alone, by the plain
sum \(S_h\), and by the joint evidence \(P_h\) alone, with the cases
outside the first tier tied.

\emph{Seed.} BEAST samples its posterior by Markov chain Monte Carlo, so
its output depends on the random seed. It was re-run on \(\Delta T\) and
flow with three further seeds.

\emph{Detector.} The changepoints of BEAST were replaced by those of
BEAST with a piecewise-constant trend and of PELT, with the same score
and ranking rule (see Changepoint detector below).

Each variation is evaluated by AP, AUC and Precision@K (P@K) on all
cases and by AUC on the season-matched subset (Table~\ref{tbl-level}).
The evaluation also reports the median number of days between the
strongest unfavourable flow change and the report in reported-fault
cases (Section~\ref{sec-timing}), and, for the ablations and seeds, the
share of the 100 highest-priority cases that are also among the 100
highest in the main analysis (top-100 overlap).

\subsection*{Parameters}\label{parameters}
\addcontentsline{toc}{subsection}{Parameters}

\begin{longtable}[]{@{}
  >{\raggedright\arraybackslash}p{(\linewidth - 10\tabcolsep) * \real{0.2871}}
  >{\centering\arraybackslash}p{(\linewidth - 10\tabcolsep) * \real{0.1485}}
  >{\centering\arraybackslash}p{(\linewidth - 10\tabcolsep) * \real{0.1485}}
  >{\centering\arraybackslash}p{(\linewidth - 10\tabcolsep) * \real{0.1485}}
  >{\centering\arraybackslash}p{(\linewidth - 10\tabcolsep) * \real{0.1485}}
  >{\centering\arraybackslash}p{(\linewidth - 10\tabcolsep) * \real{0.1188}}@{}}
\caption{Range of each metric, rounded to two decimals, when one
parameter is varied over the listed values (days for \(L\) and the
windows) with the others at the values of the main analysis, and over
the joint variation of \(q\), \(z\) and \(L\). The main analysis has AP
0.772, P@100 0.94, matched AUC 0.855 and a flow timing of 11 days.
Matched AUC is the AUC in the season-matched subset. Flow timing is the
median number of days between the strongest unfavourable flow change and
the report in reported-fault cases. ``Times as positions'' converts
BEAST's changepoint times to positions in the series (see
text).}\label{tbl-sensitivity}\tabularnewline
\toprule\noalign{}
\begin{minipage}[b]{\linewidth}\raggedright
Parameter (main analysis)
\end{minipage} & \begin{minipage}[b]{\linewidth}\centering
Values
\end{minipage} & \begin{minipage}[b]{\linewidth}\centering
AP
\end{minipage} & \begin{minipage}[b]{\linewidth}\centering
P@100
\end{minipage} & \begin{minipage}[b]{\linewidth}\centering
Matched AUC
\end{minipage} & \begin{minipage}[b]{\linewidth}\centering
Flow timing
\end{minipage} \\
\midrule\noalign{}
\endfirsthead
\toprule\noalign{}
\begin{minipage}[b]{\linewidth}\raggedright
Parameter (main analysis)
\end{minipage} & \begin{minipage}[b]{\linewidth}\centering
Values
\end{minipage} & \begin{minipage}[b]{\linewidth}\centering
AP
\end{minipage} & \begin{minipage}[b]{\linewidth}\centering
P@100
\end{minipage} & \begin{minipage}[b]{\linewidth}\centering
Matched AUC
\end{minipage} & \begin{minipage}[b]{\linewidth}\centering
Flow timing
\end{minipage} \\
\midrule\noalign{}
\endhead
\bottomrule\noalign{}
\endlastfoot
Threshold \(q\) (0.99) & 0.5--0.999 & 0.75--0.79 & 0.94 & 0.82--0.89 &
10--13 \\
Criterion \(z\) (3) & none, 1--5 & 0.70--0.77 & 0.90--0.95 & 0.75--0.86
& 10--11 \\
Reference period \(L\) (30) & 14--60 & 0.70--0.77 & 0.90--0.96 &
0.76--0.85 & 10--11 \\
Window before \(\widetilde{\tau}\) only (30) & 7--30 & 0.67--0.77 &
0.86--0.94 & 0.74--0.85 & 11 \\
Local window \(W_2\) (\(\pm2\)) & \(\pm1\)--\(\pm5\) & 0.77--0.78 &
0.93--0.98 & 0.85 & 11 \\
Refinement window (\(\pm1\)) & 0--\(\pm3\) & 0.67--0.77 & 0.88--0.94 &
0.68--0.85 & 9--13 \\
Refinement, times as positions & 0--\(\pm3\) & 0.75--0.77 & 0.92--0.94 &
0.82--0.85 & 11--13 \\
\(q\times z\times L\) jointly & 36 settings & 0.71--0.79 & 0.88--0.97 &
0.75--0.89 & 10--12 \\
\end{longtable}

\begin{figure}

\centering{

\pandocbounded{\includegraphics[keepaspectratio]{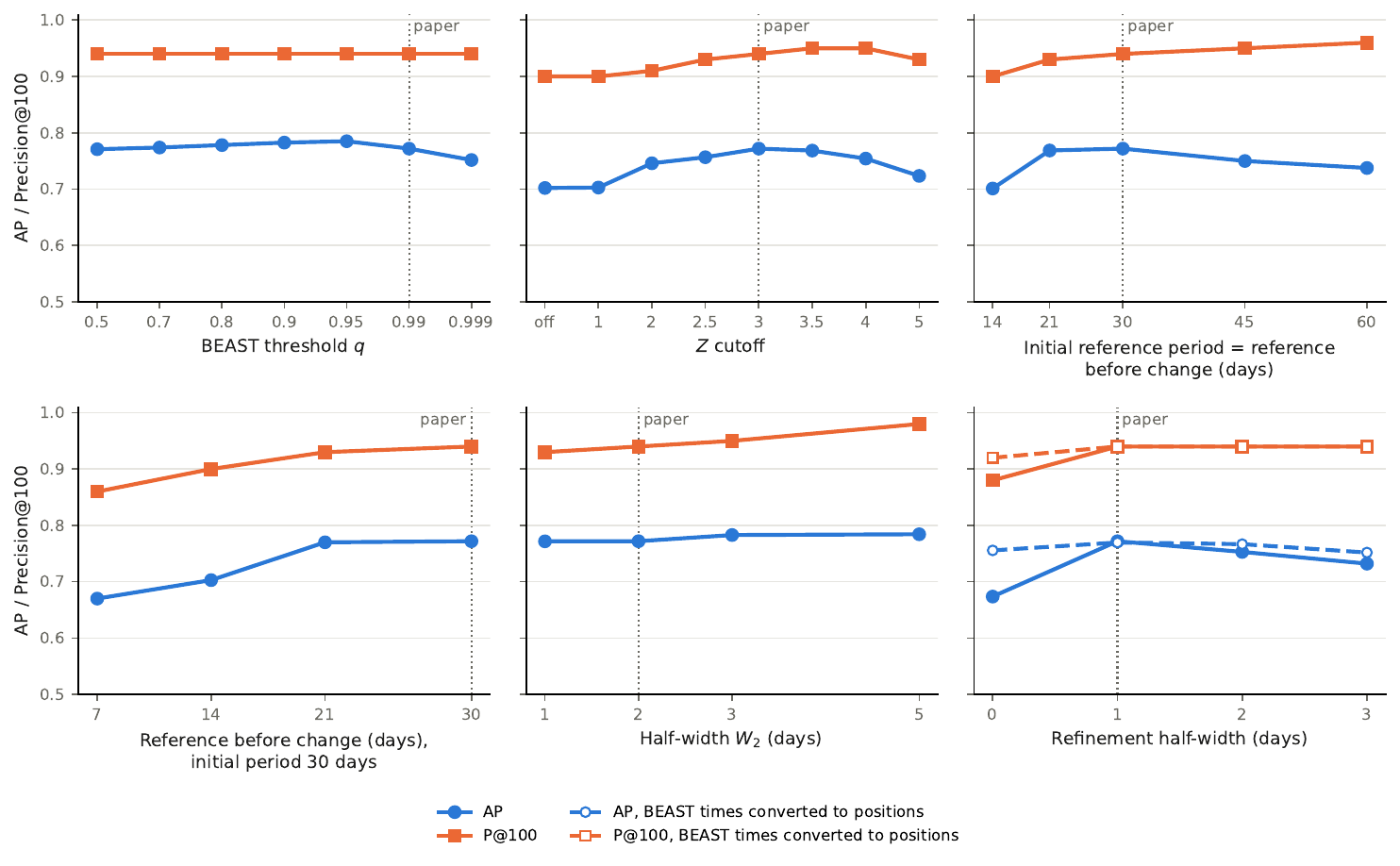}}

}

\caption{\label{fig-sensitivity}Average Precision (AP) and Precision@100
as each parameter is varied, with the others at the values of the main
analysis (dotted line). The reported-fault prevalence is 0.426. In the
refinement panel, the dashed lines convert BEAST's changepoint times to
positions in the series (see text).}

\end{figure}%

No parameter changes the conclusions (Table~\ref{tbl-sensitivity},
Figure~\ref{fig-sensitivity}). Precision among the highest-priority
cases is the most stable quantity. In every setting, at least 92\% of
the 25 highest-priority cases and at least 86\% of the 100 highest had a
reported fault. AP stayed between 0.67 and 0.79, against a prevalence of
0.426. The main analysis is not the best-performing setting: \(q=0.95\)
and a local window of \(\pm5\) days reach an AP of 0.785 and 0.784. In
the joint variation of \(q\), \(z\) and \(L\), AP ranged from 0.708 to
0.785, so no combination of these parameters changes the picture either.

Performance declines only where a setting removes what a component is
meant to do. A criterion of one standard deviation or less retains
almost every changepoint, and reference windows of 7 to 14 days give
noisy estimates of \(\sigma_{h,c}\), so that ordinary fluctuations count
as abnormal. Stricter settings, such as a higher \(q\), a higher
criterion or a longer reference period, retain fewer changepoints. The
first tier then shrinks and becomes purer, from 708 cases with 63\%
reported faults at \(q=0.5\) to 319 cases with 85\% at \(q=0.999\), and
248 with 88\% at a criterion of 5. Precision among the 100
highest-priority cases rises to as much as 0.97, while AP falls because
fewer reported-fault cases enter the first tier. This is a trade-off
between the length and the purity of the inspection list, not a change
in what the ranking finds.

The refinement window needs a remark. Rbeast counts time from 1, so a
reported changepoint, read as a position in the series, lies one day
after the jump. In the main analysis, the \(\pm1\)-day refinement
absorbs this offset and moves 84\% of the changepoints above the
threshold one day earlier. Without refinement, the score is therefore
evaluated one day after most jumps, and AP falls to 0.674. With the
changepoint times converted to positions (dashed lines in
Figure~\ref{fig-sensitivity}), AP is 0.756 without refinement and
0.752--0.770 with windows of \(\pm1\) to \(\pm3\) days. Once the times
are converted, refinement therefore changes the ranking little, and its
width matters little.

The timing result is the least sensitive. In every setting, the
strongest unfavourable flow change in reported-fault cases fell a median
of 9 to 13 days before the report, and 63--74\% of these changes fell
within the final 30 days, against about 20\% if the changes were spread
uniformly. For comparison cases, the median was at least 104 days, with
at most 10\% in the final 30 days. For \(\Delta T\), the median in
reported-fault cases ranged from 17 to 38 days. Finally, in every
setting except no refinement without the conversion of times, the AUC in
the season-matched subset exceeded the AUC on all cases. The conclusion
that the seasonal imbalance does not inflate the results therefore holds
throughout.

\subsection*{Score components and ranking
rule}\label{score-components-and-ranking-rule}
\addcontentsline{toc}{subsection}{Score components and ranking rule}

\begin{longtable}[]{@{}
  >{\raggedright\arraybackslash}p{(\linewidth - 12\tabcolsep) * \real{0.3694}}
  >{\centering\arraybackslash}p{(\linewidth - 12\tabcolsep) * \real{0.0991}}
  >{\centering\arraybackslash}p{(\linewidth - 12\tabcolsep) * \real{0.0991}}
  >{\centering\arraybackslash}p{(\linewidth - 12\tabcolsep) * \real{0.0991}}
  >{\centering\arraybackslash}p{(\linewidth - 12\tabcolsep) * \real{0.0991}}
  >{\centering\arraybackslash}p{(\linewidth - 12\tabcolsep) * \real{0.1171}}
  >{\centering\arraybackslash}p{(\linewidth - 12\tabcolsep) * \real{0.1171}}@{}}
\caption{Ablation of the score components (each replaced by 1, with the
two-tier rule) and of the ranking rule (with the full score).
``Probability only'' scores each changepoint by its signed posterior
probability. Top-100 overlap is the share of the 100 highest-priority
cases that are also among the 100 highest in the main
analysis.}\label{tbl-ablation}\tabularnewline
\toprule\noalign{}
\begin{minipage}[b]{\linewidth}\raggedright
Variant
\end{minipage} & \begin{minipage}[b]{\linewidth}\centering
AP
\end{minipage} & \begin{minipage}[b]{\linewidth}\centering
AUC
\end{minipage} & \begin{minipage}[b]{\linewidth}\centering
P@25
\end{minipage} & \begin{minipage}[b]{\linewidth}\centering
P@100
\end{minipage} & \begin{minipage}[b]{\linewidth}\centering
Matched AUC
\end{minipage} & \begin{minipage}[b]{\linewidth}\centering
Top-100 overlap
\end{minipage} \\
\midrule\noalign{}
\endfirsthead
\toprule\noalign{}
\begin{minipage}[b]{\linewidth}\raggedright
Variant
\end{minipage} & \begin{minipage}[b]{\linewidth}\centering
AP
\end{minipage} & \begin{minipage}[b]{\linewidth}\centering
AUC
\end{minipage} & \begin{minipage}[b]{\linewidth}\centering
P@25
\end{minipage} & \begin{minipage}[b]{\linewidth}\centering
P@100
\end{minipage} & \begin{minipage}[b]{\linewidth}\centering
Matched AUC
\end{minipage} & \begin{minipage}[b]{\linewidth}\centering
Top-100 overlap
\end{minipage} \\
\midrule\noalign{}
\endhead
\bottomrule\noalign{}
\endlastfoot
Main analysis & 0.772 & 0.798 & 0.96 & 0.94 & 0.855 & 1.00 \\
\emph{Score components} & & & & & & \\
Without \(p\) & 0.772 & 0.798 & 0.96 & 0.94 & 0.854 & 1.00 \\
Without \(\log(1+Z)\) & 0.758 & 0.796 & 0.92 & 0.91 & 0.846 & 0.82 \\
Without \(\log(1+J)\) & 0.728 & 0.773 & 0.88 & 0.88 & 0.815 & 0.59 \\
Without \(\log(1+C)\) & 0.772 & 0.787 & 1.00 & 0.91 & 0.838 & 0.71 \\
Without criterion \(Z>3\) & 0.702 & 0.716 & 0.96 & 0.90 & 0.754 &
0.85 \\
Probability only & 0.711 & 0.764 & 0.84 & 0.83 & 0.797 & 0.37 \\
Probability only, without criterion & 0.542 & 0.634 & 0.64 & 0.59 &
0.667 & 0.26 \\
\emph{Ranking rule} & & & & & & \\
\(\Delta T\) only, \(S_{h,\Delta T}\) & 0.699 & 0.757 & 0.84 & 0.87 &
0.815 & 0.49 \\
Flow only, \(S_{h,\mathrm{flow}}\) & 0.662 & 0.674 & 0.96 & 0.83 & 0.662
& 0.73 \\
Plain sum \(S_h\) & 0.729 & 0.764 & 0.88 & 0.88 & 0.825 & 0.57 \\
Joint evidence \(P_h\) only & 0.672 & 0.721 & 0.96 & 0.94 & 0.736 &
1.00 \\
\end{longtable}

The criterion \(Z>3\) and the jump size carry the most information
(Table~\ref{tbl-ablation}). Without them, AP falls to 0.702 and 0.728,
and the 100 highest-priority cases change substantially. The factor
\(\log(1+Z)\) adds a little beyond the criterion. The posterior
probabilities alone, signed by the direction of the change, rank the
cases poorly (AP 0.542), and with the criterion they reach 0.711, still
below the full score. The ranking is thus driven by the physically
informed post-processing rather than by the statistical confidence in a
change.

Two components contribute little in this dataset. In the main analysis,
every retained changepoint has a posterior probability above 0.99, so
the factor \(p\) is nearly constant and acts only through the threshold
\(q\). Even at \(q=0.5\) and \(q=0.9\), removing it changes AP by at
most 0.003. The consistency term \(C\) does not raise AP: without it, AP
is 0.772 at \(q=0.99\), 0.771 at \(q=0.5\) and 0.782 at \(q=0.9\). It
does, however, consistently raise precision among the 100
highest-priority cases, the part of the ranking that would be inspected
first, to 0.94 against 0.91, 0.87 and 0.90 without it. Neither component
harms the ranking, so the formulation of the main analysis is kept.

The two-tier rule achieves the highest AP and AUC of the ranking rules.
Ranking by the joint evidence alone produces the same first tier, and
therefore the same P@25 and P@100, but leaves the remaining cases
unordered. The summed score in the second tier accounts for the
difference in AP (0.672 against 0.772). \(\Delta T\) alone ranks the
cases better than flow alone, consistent with the Discussion, but both
fall short of the combination. The plain sum is dominated by
\(\Delta T\), because jumps in flow, measured in m³/h, are small.

\subsection*{BEAST seed}\label{beast-seed}
\addcontentsline{toc}{subsection}{BEAST seed}

\begin{longtable}[]{@{}
  >{\raggedright\arraybackslash}p{(\linewidth - 14\tabcolsep) * \real{0.1683}}
  >{\centering\arraybackslash}p{(\linewidth - 14\tabcolsep) * \real{0.0990}}
  >{\centering\arraybackslash}p{(\linewidth - 14\tabcolsep) * \real{0.0990}}
  >{\centering\arraybackslash}p{(\linewidth - 14\tabcolsep) * \real{0.0990}}
  >{\centering\arraybackslash}p{(\linewidth - 14\tabcolsep) * \real{0.1782}}
  >{\centering\arraybackslash}p{(\linewidth - 14\tabcolsep) * \real{0.1188}}
  >{\centering\arraybackslash}p{(\linewidth - 14\tabcolsep) * \real{0.1188}}
  >{\centering\arraybackslash}p{(\linewidth - 14\tabcolsep) * \real{0.1188}}@{}}
\caption{Results with BEAST re-run on \(\Delta T\) and flow using three
further random seeds, with the decision score and ranking rule of the
main analysis.}\label{tbl-seeds}\tabularnewline
\toprule\noalign{}
\begin{minipage}[b]{\linewidth}\raggedright
Seed
\end{minipage} & \begin{minipage}[b]{\linewidth}\centering
AP
\end{minipage} & \begin{minipage}[b]{\linewidth}\centering
P@25
\end{minipage} & \begin{minipage}[b]{\linewidth}\centering
P@100
\end{minipage} & \begin{minipage}[b]{\linewidth}\centering
First tier (fault share)
\end{minipage} & \begin{minipage}[b]{\linewidth}\centering
Matched AUC
\end{minipage} & \begin{minipage}[b]{\linewidth}\centering
Flow timing (days)
\end{minipage} & \begin{minipage}[b]{\linewidth}\centering
Top-100 overlap
\end{minipage} \\
\midrule\noalign{}
\endfirsthead
\toprule\noalign{}
\begin{minipage}[b]{\linewidth}\raggedright
Seed
\end{minipage} & \begin{minipage}[b]{\linewidth}\centering
AP
\end{minipage} & \begin{minipage}[b]{\linewidth}\centering
P@25
\end{minipage} & \begin{minipage}[b]{\linewidth}\centering
P@100
\end{minipage} & \begin{minipage}[b]{\linewidth}\centering
First tier (fault share)
\end{minipage} & \begin{minipage}[b]{\linewidth}\centering
Matched AUC
\end{minipage} & \begin{minipage}[b]{\linewidth}\centering
Flow timing (days)
\end{minipage} & \begin{minipage}[b]{\linewidth}\centering
Top-100 overlap
\end{minipage} \\
\midrule\noalign{}
\endhead
\bottomrule\noalign{}
\endlastfoot
Main analysis & 0.772 & 0.96 & 0.94 & 407 (0.813) & 0.855 & 11 & 1.00 \\
1 & 0.777 & 0.96 & 0.95 & 409 (0.814) & 0.857 & 11 & 0.98 \\
2 & 0.773 & 0.96 & 0.94 & 401 (0.815) & 0.856 & 11 & 0.96 \\
3 & 0.774 & 0.96 & 0.94 & 408 (0.806) & 0.864 & 11 & 0.96 \\
\end{longtable}

The seed of the BEAST sampler has almost no effect
(Table~\ref{tbl-seeds}). Across seeds, the number of changepoints above
\(q\) changed by less than 1.5\%. AP changed by at most 0.005, and P@25,
the size and purity of the first tier, and the median flow timing were
essentially unchanged. The Spearman rank correlation with the ranking of
the main analysis was 0.95 to 0.96, and 96 to 98 of the 100
highest-priority cases were the same. The ablations are equally stable.
For every seed, removing the criterion or the jump-size term lowered AP
by 0.04 to 0.07, removing \(C\) left AP within 0.004 of the full score
while lowering P@100 by 0.01 to 0.03, and the two-tier rule had the
highest AP of the ranking rules.

\subsection*{Changepoint detector}\label{changepoint-detector}
\addcontentsline{toc}{subsection}{Changepoint detector}

The decision score takes the changepoints of a detector as input, so its
value should not depend on BEAST. Two alternatives were run on
\(\Delta T\) and flow and passed through the same score and ranking
rule. The first is BEAST with a piecewise-constant trend, which models
shifts in the level, instead of the piecewise-linear trend of the main
analysis. The second is PELT (19), as implemented in the ruptures
package (25), with either a mean-shift cost or a continuous
piecewise-linear cost. PELT returns no probabilities, so \(p\) was set
to 1. Its penalty \(c\,\hat\sigma^2\log n\), with \(\hat\sigma\) a
robust estimate of the noise level from the first differences and \(n\)
the window length, was varied over \(c\in\{2,4,8,\dots,256\}\). Choosing
PELT's penalty on the evaluation data favours PELT, whereas BEAST's
threshold was fixed in advance. The \(\pm1\)-day refinement was applied
to all detectors.

\begin{longtable}[]{@{}
  >{\raggedright\arraybackslash}p{(\linewidth - 10\tabcolsep) * \real{0.2178}}
  >{\centering\arraybackslash}p{(\linewidth - 10\tabcolsep) * \real{0.1485}}
  >{\centering\arraybackslash}p{(\linewidth - 10\tabcolsep) * \real{0.2178}}
  >{\centering\arraybackslash}p{(\linewidth - 10\tabcolsep) * \real{0.1683}}
  >{\centering\arraybackslash}p{(\linewidth - 10\tabcolsep) * \real{0.1485}}
  >{\centering\arraybackslash}p{(\linewidth - 10\tabcolsep) * \real{0.0990}}@{}}
\caption{The decision score and ranking rule applied to the changepoints
of different detectors. For each detector, the first row is a single
setting, chosen for PELT to give about as many changepoints as BEAST in
the main analysis, and the second row the range over all settings. Flow
timing is the median number of days between the strongest unfavourable
flow change and the report in reported-fault
cases.}\label{tbl-detectors}\tabularnewline
\toprule\noalign{}
\begin{minipage}[b]{\linewidth}\raggedright
Detector and model
\end{minipage} & \begin{minipage}[b]{\linewidth}\centering
Setting
\end{minipage} & \begin{minipage}[b]{\linewidth}\centering
Changepoints per window (\(\Delta T\) / flow)
\end{minipage} & \begin{minipage}[b]{\linewidth}\centering
AP
\end{minipage} & \begin{minipage}[b]{\linewidth}\centering
P@100
\end{minipage} & \begin{minipage}[b]{\linewidth}\centering
Flow timing
\end{minipage} \\
\midrule\noalign{}
\endfirsthead
\toprule\noalign{}
\begin{minipage}[b]{\linewidth}\raggedright
Detector and model
\end{minipage} & \begin{minipage}[b]{\linewidth}\centering
Setting
\end{minipage} & \begin{minipage}[b]{\linewidth}\centering
Changepoints per window (\(\Delta T\) / flow)
\end{minipage} & \begin{minipage}[b]{\linewidth}\centering
AP
\end{minipage} & \begin{minipage}[b]{\linewidth}\centering
P@100
\end{minipage} & \begin{minipage}[b]{\linewidth}\centering
Flow timing
\end{minipage} \\
\midrule\noalign{}
\endhead
\bottomrule\noalign{}
\endlastfoot
BEAST, piecewise-linear trend (main analysis) & \(q=0.99\) & 2.5 / 1.8 &
0.772 & 0.94 & 11 \\
& \(q\): 0.5--0.99 & 2.5--7.1 / 1.8--4.9 & 0.771--0.783 & 0.94 &
11--13 \\
BEAST, piecewise-constant trend & \(q=0.99\) & 2.7 / 2.0 & 0.790 & 0.94
& 11 \\
& \(q\): 0.5--0.99 & 2.7--7.3 / 2.0--5.4 & 0.790--0.800 & 0.94 &
11--14 \\
PELT, mean shifts & \(c=64\) & 2.6 / 1.1 & 0.794 & 0.95 & 14 \\
& \(c\): 2--256 & 1.1--20.5 / 0.4--10.2 & 0.733--0.807 & 0.90--0.98 &
12--17 \\
PELT, continuous piecewise-linear & \(c=128\) & 2.8 / 1.4 & 0.638 & 0.92
& 13 \\
& \(c\): 2--256 & 1.9--24.3 / 0.9--14.3 & 0.627--0.784 & 0.92--0.97 &
12--15 \\
\end{longtable}

The decision score works with every detector
(Table~\ref{tbl-detectors}). Across all detectors and settings, at least
92\% of the 25 highest-priority cases and at least 90\% of the 100
highest were associated with a reported fault, and the strongest
unfavourable flow change in reported-fault cases fell a median of 11 to
17 days before the report. The ranking and timing results are therefore
largely properties of the score rather than of BEAST.

The candidate model matters more than the detector. Models of shifts in
the level, which is what \(Z\), \(J\) and \(C\) measure, ranked the
cases better than models of linear trends. BEAST's AP rose from
0.771--0.783 to 0.790--0.800 with a constant trend, and PELT reached
0.733--0.807 with a mean-shift cost against 0.627--0.784 with a linear
cost. For a given model, BEAST was the more reliable detector. With
linear trends, PELT fell to an AP of 0.638--0.679 at a number of
changepoints comparable to BEAST's. With mean shifts, PELT's AP varied
with the penalty from 0.733 to 0.807, whereas BEAST's varied by at most
0.012 over \(q\) from 0.5 to 0.99. BEAST also dated the changes more
sharply: at \(q=0.99\), the median flow timing was 11 days for both
BEAST models against 12 to 17 days for PELT, and 69--70\% of the changes
fell within the final 30 days against 62--67\%.

The main analysis was not changed to the piecewise-constant model,
because that choice would have been made after seeing the results. Its
gain of about 0.02 in AP is reported here as a direction for refinement.

In summary, none of the parameters, components, seeds or detectors
examined changes the qualitative conclusions of the paper. The choices
of the main analysis lie on a broad plateau of comparable performance.
The elements that matter are those with a clear physical role, and the
timing of the detected flow changes relative to the report is
essentially invariant. These checks were made on the data on which the
score was developed and do not replace validation on an independent
district-heating system.

\end{document}